\documentclass[11pt]{article}
\usepackage{times}
\usepackage[numbers,sort&compress]{natbib}

\usepackage[margin=1in]{geometry}

\usepackage[compact]{titlesec}  
\titlespacing{\section}{0pt}{0pt}{0pt}
\usepackage{comment}

\begin{document}

\title{
Arti-``fickle" Intelligence: Using LLMs as a Tool for Inference in the Political and Social Sciences
}

\author{Lisa P. Argyle, Ethan C. Busby, Joshua R. Gubler, Bryce Hepner, Alex Lyman, David Wingate}

\maketitle

\begin{abstract}
\noindent Generative large language models (LLMs) are incredibly useful, versatile, and promising tools. 
However, they will be of most use to political and social science researchers when they are used in a way that advances understanding about real human behaviors and concerns. To promote the scientific use of LLMs, we suggest that researchers in the political and social sciences need to remain focused on the scientific goal of inference. To this end, we discuss the challenges and opportunities related to scientific inference with LLMs, using validation of model output as an illustrative case for discussion. We propose a set of guidelines related to establishing the failure and success of LLMs when completing particular tasks, and discuss how we can make inferences from these observations. We conclude with a discussion of how this refocus will improve the accumulation of shared scientific knowledge about these tools and their uses in the social sciences. 
\end{abstract}

\parindent 0.0in
\parskip 0.15in
\thispagestyle{empty}

%
%
\section{Introduction}

The political and social sciences have quickly adopted generative AI into virtually every aspect of the research process.
Large language models (LLMs) are used for idea and theory generation \citep{meincke2024using, si2024}, as automated research assistants \citep{schmidgall2025agentlaboratoryusingllm, agarwal2024llms}, for writing code \citep{LLMsForScience}, for annotating and classifying text and image data \citep{tornberg2024large, ornstein2023train, heseltine2024large, wang2024llms, rytting2023}, for producing and administering experimental manipulations \citep{velez2023confronting, argyle2023ai, tessler2024, hackenburg_margetts_2023}, as the object of study \citep{rozado2023political, park2023generative, palmer2023large}, for simulation of subject data \citep{argyle2022does, tornberg2023simulating, sreedhar2025simulating, bisbee2023artificially, ashokkumar2024predicting}, and even as direct partners in writing up results \citep{thorp2023chatgpt, van2024authorship}. 
A broad search through published work in the social sciences suggests that the number of published articles using generative AI tools increased by a factor of 500 percent from 2023 to 2024, with no signs of slowing.\footnote{Calculated from a citation search of published articles in the social sciences on the Web of Science for the key terms ``large language model," ``generative AI," and ``artificial intelligence."}
In political science, the 2023 annual meeting of the American Political Science Association included 10 research papers using generative AI and one ``breaking news" panel on large language models (LLMs) in political science; the same conference in 2024 included a full-day pre-conference on LLMs and over 100 papers making some use of generative AI and LLMs.\footnote{These counts come from names and brief descriptions of panels and papers in the conference program.} 

Like others, we are enthusiastic about the potential contributions of LLMs  across the social sciences \citep{bail2024can, grossmann2023ai, xu2024ai};  
LLMs have enabled our team to explore a range of interesting and complex questions in ways that would not have been otherwise possible \citep{argyle2022does,argyle2023ai, argyle2024, lyman2025smr}. 
%
LLMs are unlike other scientific tools: they are neither a statistical model with carefully bounded properties, nor a machine learning algorithm with well-defined inputs, outputs and optimization objectives \citep{mccoy2024embers}. They are ``programmed'' with prompts that are remarkably fragile \citep{liu2024poliprompt, atreja2024prompt, zhuo-etal-2024-prosa}, often trained on an opaque mix of data and aligned to secret standards \citep{spirling23, ollion2024dangers}, and their human-like outputs are presented so naturally and confidently that it's all too easy to forget they need to be validated \citep{jazwinska2025ai}. 

The rise of LLMs is more like the advent of the internet, which  transformed virtually all aspects of social science research, than it is like other social scientific tools.
As journal repositories moved online, the internet opened new possibilities for efficiently accumulating and summarizing existing scholarly research. 
Access to digitized administrative data sources and diverse survey subject pools became faster, easier, and less costly than ever before. 
Internet searches changed the way students learn. 
Social media altered they way people interacted with and related to each other, opening new avenues for observation of and data collection about human behavior. 

LLMs are again reshaping each of these domains, including summarization and synthesis of scholarly research \citep{briggs2025}, 
simulation of human survey or experimental data \citep{dillion2023can, argyle2022does, bisbee2023artificially, ashokkumar2024predicting}, and interactions with survey research subjects \citep{argyle2023ai, velez2024, tessler2024}. 
LLMs facilitate administrative record linkages \citep{Li2023, Fernandez2023}, change the way students learn \citep{Xiao2023,lyu2024, milano2023}, and change the way individuals interact \citep{yakura2024empiricalevidencelargelanguage, Hohenstein2023}. We therefore consider it important to discuss how these tools can be best used to not simply generate knowledge, but to \textit{generate the type of cumulative scientific inference that is the goal of social scientific research.}


In the sections that follow, we begin with a brief discussion of what we mean by scientific knowledge, with a focus on the goal of scientific inference. 
We then discuss the implications of this goal for the use of LLMs in the social sciences, focusing on the importance of clearly mapping LLM outputs to targets of inference, and on the importance of systematic validation of LLM uses, including reporting and interpreting LLM success and failure modes for the range of potential social science tasks. 

While we feel that clearly establishing guidelines for inference and validation in LLM use will greatly facilitate the cumulation of community-wide scientific knowledge, we also note that these are not the only facets of the LLM research process that need further discussing in the pursuit of science: similar conversations should happen around other aspects of LLMs in the research process, including but not limited to ethics, replication and transparency, and governance and regulation. 
Our contribution here, then, is an illustrative case of the kinds of complex considerations at play regarding the scientific use of LLMs and how social scientists can be more thoughtful about integrating LLMs into their scientific pursuits. 

%
%


\section{Social Scientific Inference Using LLMs}


Although conceptions of what exactly constitutes science vary, ``positive'' social scientists generally agree that the fundamental goal of science is to infer descriptive or causal conclusions that go beyond a specific set of observed data \citep{kkv2021}. In this view, scientific knowledge is not just knowing ``X caused Y in this particular instance," but instead identifying the conditions under which, and the confidence by which we can expect X to cause Y \textit{generally}. Thus, theory is scientific knowledge, but facts (while helpful in the pursuit of scientific knowledge) are not. As Lakatos notes, ``A given fact is explained \textit{scientifically} only if a new fact is also explained with it'' \citep[emphasis added, p. 119]{lakatos1970_2} -- in other words, only if there is a theory about an underlying, consistent pattern that explains more than one instance.  

From this premise, we argue that LLM research that notes that a particular model can or cannot do something -- meaning it establishes a ``fact'' -- without careful thought and discussion of ``why'' and ``under what conditions,'' does not generate scientific knowledge. While such factual knowledge can be useful for a variety of reasons, it tells the scientific community little unless it can generalize to other research and/or to the broader use of LLMs. Given these basic insights, we suggest that at a minimum, scientific research using LLMs requires the following:
\begin{enumerate}
    \item Clearly defining the \textit{target} of inference
    \item Determining when and under what conditions a particular inference will be \textit{valid}
\end{enumerate}

To meaningfully contribute in a scientific way, social scientists using LLMs must first make the target of their inference clear. By this we mean that social scientists ought to have in mind a broader theory or conclusion that would be supported by establishing specific facts using the LLMs. For example, social scientists know that when running a randomized experiment, it is vital to identify the type of thing their treatment represents ``in the real world.'' Likewise, when using observational data or case studies, social scientists clearly define how their particular measures and observations can be used to infer to outcomes and patterns in the larger social and political world. 

Clearly identifying the target of inference in our studies is scientifically crucial for multiple reasons. Most importantly, it helps other scientists understand what the results mean and where they can be expected to apply. In the case of LLM-enabled research, identifying the target of inference links a given use of an LLM to a larger literature, set of theories, and core hypotheses, or to the creation of new literatures and hypotheses. It also naturally sets expectations for the types of validation tasks necessary to trust the results. 
Without this kind of clarity, the scientific value of an LLM demonstration will remain at least partially unclear if not impossible to determine. 



For example, a growing number of scholars use LLMs to examine how AI agents produce or respond to treatments in randomized experiments or other interactive settings \citep{Cui2024, horton2023large, Lippert2024, park2023generative, tornberg2023simulating, lyman2025smr, ashokkumar2024predicting, palmer2023large, hackenburg2025scaling}. In these types of studies, the target of inference is related to the LLM itself -- does an LLM have the capacity to complete a particular, desired task, with implications for future research? Even within this scope, these studies might have any number of important inferential targets: the ability of the models to produce treatments or interactions in a consistent way, the ability of the models to generate results similar to the human-based studies to which they are compared, or the abilities of the models to tell us something novel about current or potential human behavior, to name a few.
If the authors do not clearly state the inferential purpose of the demonstration,\footnote{We intend this as a conditional statement, and not an indictment of any of these particular studies.} 
then the scientific community will be unable to apply the accumulated facts to a meaningful inference or accurately evaluate what the demonstration means for future uses of the tools. 

As a further example, another set of studies have significant overlap in content and design, but feature LLMs used to create and administer some kind of intervention or treatment to human subjects \citep{argyle2023ai, tessler2024, hackenburg_margetts_2023, velez2023confronting}. In these cases, the target of inference is not the LLM itself,  but rather the experience or causal process the LLM helps facilitate. As such, study validation will look different, with validation focused on whether the LLM produced a treatment that was consistent with the researcher's intended manipulation or the real-world corollary. Thus, the researcher needs to establish validity of the LLM's completion of the task in order to be confident in inferences made about theories of human behavior, which are directly tested with human subjects. 

It is therefore important to make the inferential target clear in each type of study, as it determines both the kind of validation necessary for the LLM and the contribution to cumulative scientific knowledge made by the specific facts of the research data collection. 
While the target of inference is often implicit in much social and political science research, we suggest that LLM-enabled research will require explicit focus on inference in order to facilitate community-wide attention to model validation procedures -- in particular to the pursuit of systematic learning from model successes and failures.

\section{Making Valid Inferences when LLMs Fail}

At times, LLMs generate such compelling results 
that it is easy to forget the need for validation. At other times, despite many attempts, LLMs seem to fail at the tasks we assign. If approached scientifically, both cases, and all others in between, can help in the cumulation of general scientific knowledge. In the following, we highlight the need to evaluate both model ``failures'' and ``successes''  and suggest some guidelines for each.

If an LLM fails at our first (or second, or fiftieth) attempt to complete a research-related task, what should we make of that failure?
Even if an LLM is fickle, conditional, limited, or unstable, a single \textit{successful} proof-of-concept is enough to demonstrate a potential capacity.
But the converse is not true: a single \textit{failure} is not sufficient to make sweeping claims about lack of capacity. 
How, then, should social scientists think about LLM failures and their role in building collective knowledge?

\subsection{Potential Root Causes of LLM Failures}

Ideally, researchers will identify the root cause of failure, which enables both documentation of what can accurately be inferred from the failure and also suggests the scope of potential solutions to remedy a failure. 
Here we describe four potential root causes of LLM task failure and suggest that when researchers grapple with failure in an LLM, 
they work through a list of these or similar possibilities to better understand the issue. By doing so and publicly sharing their conclusions, researchers can make valid inferences about the failure, allowing for cumulative scientific knowledge building.

\textbf{The Task is Beyond the Capacity of Any Language Model:} 
LLMs are highly versatile, and the full scope of their capacities is still being mapped across a wide variety of academic and industrial use-cases. 
However, we do not expect that they are capable of completing every task, and it is possible that an LLM fails because the researcher is asking it to do something that is fundamentally beyond what it can provide \citep{mccoy2024embers}. 
For example, current LLMs struggle with tasks that require deep subject matter expertise \citep{szymanski2025limitations}, genuine human-like emotion \citep{huang2024apathetic}, or complex reasoning and logic \citep{amirizaniani2024can, valmeekam2022large}.
Sometimes even seemingly simple tasks are beyond the capabilities of flagship LLMs, such as public failure of multiple LLMs to count the number of ``R''s in the word ``strawberry'' \citep{eaton2024}.

For failures of this type, the appropriate response is to seek an entirely different tool to complete the task, at least until new AI tools with new potentials are developed.

\textbf{The Task is Beyond the Capacity of the Particular Language Model Used:} 
The capability of any particular language model depends on the training data and architecture upon which it is built. 
%
For example, an LLM may be unable to provide high-quality translation into a language which is only minimally or not at all present in its training data; however, if the model were re-trained on a corpus that included sufficient text in the new language, it could perform better \citep{lu2024llamax}. 
Similarly, larger models are generally capable of more things than smaller models \citep{hackenburg2025scaling,kaplan2020scalinglawsneurallanguage}, and LLMs in the same family tend to perform comparably to each other, and differently from LLMs generated by other providers. 

For all of these reasons, failure might not occur because LLMs as a class of tool cannot complete the task, but rather because the chosen LLM used is insufficient.  
If so, the appropriate response may be to replicate the task on one or more different models, across sizes and families, to identify and select the best-performing model for the task.     

\textbf{The Language Model has the Capacity to Complete the Task, but Alignment Interferes:} 
There are some tasks an LLM can do, but will resist doing or will do in a distorted way, because of post-training alignment. 
Alignment serves two distinct purposes: first, it enhances LLMs' capacity to follow instructions (and thereby to be more helpful), and second, it reduces undesirable behaviors (by, for example, refusing to answer queries deemed harmful).
Following instructions is unquestionably useful, and there are many situations when reducing harmful output is desirable. However, for other social science research applications, refusals or other distortions to the LLM output can interfere with task completion. 
Alignment often causes models to \textit{over-refuse} \citep{cui2024orbenchoverrefusalbenchmarklarge}, meaning it refuses to complete a benign request miscategorized as harmful.
For example, multiple researchers have documented LLMs failing or refusing to complete a task because of the presence of a racial stereotype in the prompt, even when the stereotype was irrelevant to the task \citep{lyman2025smr, tekgurler2025llms}. Likewise, aligned models tend to have lower variance in their output \citep{kirk2024understanding}, and the ``helpful assistant'' persona aligned into LLM output may interfere with the form or tone of a desired response \citep{lyman2025smr}. 
%

If alignment causes a task failure, there are a few potential responses the researcher might take. 
In some cases, careful prompt engineering to escape alignment (often termed ``jailbreaking'') is sufficient.
Or, since different models from different companies are aligned with different procedures, choosing a different model may yield different behavior.
Finally, researchers could fine-tune a language model themselves to better perform the desired task.
We note, however, that fine-tuning is the most resource-intensive option, and sometimes results in reduced capability or ``catastrophic forgetting" of other capacities \citep{li-etal-2024-revisiting}.

\textbf{The Language Model has the Capacity to Complete the Task, but the Researcher is Not Efficiently Accessing or Extracting the Requisite Performance:} 
Good prompting of a language model is critical. 
Even small variations in a prompt, including the order in which text is given or wording choices that seem semantically equivalent, can have a substantial impact on model output and performance \cite{pezeshkpour-hruschka-2024-large}.
However, prompting best practices are still emerging, and it is not immediately apparent by looking at a prompt whether it is ``good."
Nor is it often possible to create a ``good'' prompt without conducting substantial testing for a specific task and with a specific model. 
Specific prompts or prompting tricks often do not translate well from one model or task domain to another \citep{zhuo-etal-2024-prosa}. 

Here, the solution is often additional prompt engineering and careful testing.
Problems of response structure, patterns of undesired responses, and misunderstanding of core concepts or instructions are common problems that can often be resolved by refining prompts.

\subsection{Interpreting LLM Failures}




Often, the root cause of LLM failure is not readily apparent. In these cases, we caution against two different inferential failures: overinterpretation and underinterpretation. 
Overinterpretation of model failure -- by which we mean taking a low-level failure, such as a poor prompt, as evidence of a high-level failure, such as a fundamental incapacity in LLMs -- poses a threat to research development.
Social scientists should be careful, particularly when publishing research questioning the potential or capacity of LLMs not to claim that a failure of a particular model, with particular settings and a particular prompt, is evidence of a general failing of LLMs as a tool.
To be properly justified, such a claim requires the researcher to convincingly rule out other potential failure modes. 

Likewise, underinterpretation of model failure -- by which we mean taking a high-level failure, such as a model that is improperly trained or mis-aligned for a task, as a lower-level problem, such as poor prompting -- may lead researchers to expend too many resources in search of an improvement of performance that may never come. 
We view both over- and under- interpretation as problems for the research community to carefully avoid in producing and discussing scholarship around the failure of LLMs.

Both dangers suggest it would be fruitful to develop reasonable tests to identify which of the failure modes is at play, likely in deep collaboration between social and computer scientists.

%
%
\subsection{Recommendations: Detailed and Transparent Reporting of Failures}

To learn from LLM failures, our community will need to avoid the ``file drawer problem.'' It is remarkably valuable for scholars to publicize their failed efforts, because such efforts reduce unintentional duplication of work across teams, and the associated wasted resources. 
As such, we urge editors and reviewers to reward researchers for reporting early-stage challenges, documenting things like changes to LLM prompts, models, or hyperparameters to maximize the validity of outputs. 

Thus, our recommendation for researchers who encounter LLM failures at any stage of the process is to robustly report all the important details. At a minimum, standards of reporting should include: 
\begin{enumerate}
    \item The particular LLM used
    \item The date on which the call was run
    \item The mode in which the LLM was accessed, such as API vs. local hosting, and any software packages used in the interface
    \item All hyperparameters of the model, including temperature, token limits, etc.
    \item The complete text of the prompt(s)
\end{enumerate}
We suggest this information appear in the main text of the article, with an appendix including additional details such as prompt iterations in the development process and descriptions of other models, parameters, and prompting approaches that were tested. 
Providing this information will help reviewers reduce their demands for additional variations and iterations in their reviews and also improve communication about best practices, which will improve efficiency for future researchers. 

Only by working through different modes of failure -- in a process similar to ruling out competing theoretical explanations for an observed results -- can social scientists make make significant collective scientific gains from failure in their work with LLMs.

%
%

\section{Validating LLM Successes to Improve Inference}

LLMs are powerful tools for social scientists because there are many things they do well. How can we determine when an LLM is doing a task \textit{well enough} for us to be confident in inferences that rest on LLM-enabled procedures? We now turn our attention to how scholars can contribute to general scientific knowledge as they systematically validate particular instances of model ``success'' in their research. We first highlight some challenges to benchmarking LLM task success. We then provide suggestions for how social scientists can rigorously and scientifically document 
the instances when LLMs generate useful, reliable results. 

\subsection{Challenges to Benchmarking LLM Success}

One definition of the content of science itself is the procedures and process by which new knowledge is accumulated \citep{kkv2021}. Because LLMs are a new tool, we do not yet have established standards for the procedures and processes of working with LLMs that generate valid inferences. Additionally, the sheer range of tasks to which an LLM can be applied means that there is no single benchmark or statistical test that can be universally applied. This creates a number of challenges related to the use of LLMs, even when we believe they are succeeding at the desired task. 

\textbf{Success is Often Multifaceted and Hard to Define:} 
Consider a scholar using an LLM to identify and summarize potential sources for a literature review. Success would likely mean establishing that the LLM is retrieving sources that are real (not hallucinated), that cover the full range of relevant scholarship in the given domain, and that any summaries or synthesis reflect the actual content of each individual study. 
While some of these factors are easy to verify, others (e.g. does the body of work reflect the complete and current state of the field?) require a high level of expertise, and even comparably trained experts might disagree. 
In some ways, the allure of LLMs is that they appear to accurately perform such complex, multifaceted assignments. 
However, validating this promise presents high hurdles and likely requires multiple metrics.

\textbf{Even Highly Related Task Domains May Have Different Indicators of Success:}
Different research teams and traditions may have different indicators of success, even in highly related task domains.
To illustrate this, consider the case of silicon sampling, a task where a scholar uses a language model to simulate human subjects to estimate political attitudes and public opinion.\footnote{Wang et al. \citep{wang2025limits} provide a more comprehensive summary of the current state of literature in this area, including a robust discussion of how to interpret failure.} 
Depending on the particular target of inference, a researcher might be interested in replicating the variance of a distribution of human responses, replicating an aggregated average,  simulating a relationship between a stimulus and outcome, or  in a precise prediction of attitudes from a particular demographic subgroup or even a specific individual.
The metrics of ``success'' implied by these objectives will differ significantly one from another, and it is unlikely that success on one will imply success on the others.

For example, Argyle et al. \citep{argyle2022does} focused on replicating mean values and relationships between demographics and attitudes for a human sample in the US. 
By contrast, Santurkar et al. \citep{santurkar2023whose} focused more on carefully replicating human attitudes for particular subgroups, finding that some groups were better represented than others by LLM output. 
Bisbee et al. \citep{bisbee2023artificially} and Boelaert et al. \citep{boelaert2024machine} focused on distribution variance in LLM vs. human attitudes. 
Kim and Lee \citep{kim2023ai} focused on providing accurate estimates for narrow subpopulations or single individuals, making a case that fine-tuning can help LLMs provide more accurate estimates.  %

How should a scholar make sense of these competing results? 
Are LLMs good or bad at the task of simulating human subjects, and what standards should one use to determine the answer? 
This example illustrates a common phenomenon in LLM-based research:
LLMs can perform very well on different tasks as evaluated by certain metrics.
LLMs can also perform poorly on other (potentially similar) tasks or the same task when evaluated using different metrics. One key to defining success is clearly identifying and justifying the target of inference and the metrics by which that inference will be validated beforehand.

\textbf{Success is Non-Transitive (at least for now):} 
At the present, success in LLMs appears to be non-transitive. This means that scholars cannot treat the validation of LLMs in the same way that they approach the adoption of other machine learning or statistical tools. 
It is common practice in political science, for example, for ``methods'' scholars to publish new methodological approaches, which typically provide formal mathematical proofs and/or applied demonstrations of the qualities and comparative advantages of a new tool.
This allows future researchers to pick up the tool, cite the appropriate scholarship justifying its use in this case, and apply it without further theoretical or empirical validation. 

LLMs, by contrast, are better understood as having varied, contingent, and multi-dimensional capacities. 
This implies that scholarly demonstration of a singular specific capacity (or lack thereof) does not imply similar performance on any other capacity, even those that seem reasonably comparable. 
As such, scholars cannot assume a model's competence at a needed task based on its global reputation, nor can they unquestioningly use previous research or existing benchmarks as a reliable heuristic for success at their particular task. Likewise, the instability of prompting \citep{zhuo-etal-2024-prosa} means that wholesale adopting a prompt from published work and applying it to a new project, with a different model, architecture, or task output, and expecting optimal performance is unlikely to produce the desired results.

\textbf{Success is Model-dependent, and Models Change Quickly:} The rapid advancement of LLMs means that particular models quickly become outdated or even unavailable for additional use. 
This has been of particular concern for social scientists using proprietary API-access models \citep{spirling23, ollion2024dangers}, which are changed or deprecated at the will of their providers. 
Such changes can make replication difficult or impossible. 
However, freezing all social science research into a particular set of saved and stable models poses its own challenge for the demonstration and evaluation of success at new tasks, where newer models may have new training and capacity better aligned to social science tasks. 
Additionally, an outcome that includes thousands of sets of frozen model weights that are stored by research teams but never in practice used again, poses its own concerns regarding resources, environmental impacts, and the practical meaning and role of replication.

%


\subsection{Recommendations: Validation as a Bespoke and Transparent Process}

This list of challenges to identifying the degree to which a model successfully completes research tasks is not exhaustive. It's likely that new advances will raise other concerns and render some of these obsolete.
In the meantime, how should scholars think about best practices for identifying and learning from LLM successes in our research?

Our primary recommendation is that validation should happen in \emph{every} social science use of a generative language model. 
The nature and extent of such validation might vary, and could range from qualitative expert evaluation to systematic quantitative benchmarking.
We therefore encourage scholars to incorporate careful thinking about appropriate validation as an early and intentional step in the research process, and we encourage reviewers to be attentive and rigorous in pushing researchers to justify their validation steps. 



Consistent with widespread norms of open science and pre-registration in political science and other social science disciplines, we propose a framework for thinking about LLM validation that can be easily incorporated into, but is not necessarily dependent upon, those existing practices. 
We see two equally workable approaches: 

\textbf{Approach 1: Conducting and Reporting Validation As Part of Study Pre-registration} 
This first approach asks scholars to conduct all validation of their LLM prior to completing and submitting study pre-registration. 
In this approach, authors describe their LLM validation process, including how they selected a particular model, the parameters of that model, and specific model prompts, and then share the results of their validation process. 
Such a pre-registration would be filed immediately prior to study launch.
This approach is best for projects where the uses of an LLM are relatively discreet in time and scope, as it requires significant justification for any deviations that happen during the study. 

\textbf{Approach 2: Pre-registering Validation Standards and Procedures} 
A second approach would ask scholars to clearly describe the specific target benchmarks or standards that, when achieved by the LLM, will be sufficient for the research to proceed with confidence. 
In this case, the researchers do not need to know at the time of pre-registration which exact prompt or model they will use. Instead, they simply identify a clear and specific standard -- or set of standards -- to identify the validty of LLM output. 
In the write-up of the study, scholars would clearly demonstrate that the eventual combination of model, parameters, and prompts employed in the study achieve these standards. 
This approach provides additional flexibility for projects with interconnected tasks or a long time horizon (in which models might change or update), while not sacrificing transparency in the research process.

These approaches are not necessarily exclusive, and a single project could take both approaches. 
%
We also note that proper clarity and transparency about the use of an LLM can be incorporated in any manuscript, whether or not it was pre-registered. 
We encourage scholars to continue to be transparent about reporting the complete use of the LLM and their validation process and standards.


One benefit to this framework is that it means every manuscript should not be required to demonstrate performance across all models or prompt variations. 
Depending on the target of inference, some studies might require extensive and systematic model output comparisons. 
In other cases, a model that performs a single task well enough for the researchers to proceed is sufficient, without requiring a complete mapping of the full potential model and prompt space to demonstrate it is the ``best" instantiation. 

%
%
\section{Summary Guidelines for the Scientific Use of LLMs}


In the course of this commentary, we have suggested several guidelines to improve researchers' abilities to build cumulative scientific knowledge using LLMs in the political and social sciences.
Because LLMs are multipurpose, fluid, and sometimes fickle, we propose general guidelines instead of specific methodological practices or statistical tests. 
We intend these guidelines to be flexible enough to accommodate a tool that will undoubtedly change, but concrete enough to provide structure to a growing research community. 
In sum, we recommend that:

\textbf{Prior to the start of a study using AI}, researchers should:
    \begin{enumerate}
        \item Clearly define the target of inference. Researchers should clearly state what they are hoping to make inferences about and what element of the LLM-empowered research design corresponds to or enables inference about that target.
        \item Identify one or more standards by which they will determine whether an LLM has successfully completed a needed task. These standards should be specific, tailored to the particular use case, and comprehensive. 
        \item Provide pre-registration showing either a) evidence that a particular combination of model, prompt, and parameters already sufficiently meets the identified standard of success, or  b) the standard and process by which the model, prompt, and other parameters will be selected and deemed sufficient for inference. 
    \end{enumerate}

\textbf{In reporting post-study data}, researchers should:
    \begin{enumerate}
    \setcounter{enumi}{3}
        \item Report the inferential target early in the paper so that this focus and objectives are clear to readers, reviewers, and other social scientists.
        \item Provide confirmatory evidence that the LLM faithfully completed the particular study task(s) as deployed.
        \item Report, ideally in the main text of the manuscript, the relevant details of LLM use, including exact models, dates of calls, mode of access, hyperparameters, and prompts.
        \item Report, in supplemental information, a summary of significant explorations of other models, prompts, settings, or approaches taken in the process of seeking task success.
    \end{enumerate}

We believe these standards fit naturally and reasonably into the existing social science research pipeline and that they provide a number of benefits for the broader research community, including transparency for stakeholders and a means to engage in community-wide, scientific knowledge building.


\section{LLMs and the Future of Inferential Social Science}

For the sake of space and clarity, we have focused primarily on suggestions to refocus LLM studies on inference and validation. 
These topics are an essential first step towards knowledge growth.
However, this focus has left aside many other important considerations that arise when using an LLM in social science research, including issues of regulation and governance, ethics, and transparency and reproducibility. 
These are not trivial matters nor are there consensus views established on these topics.
To use LLMs for quality science, political and social scientists should be deeply engaged in conversations on these and other fronts. 
Indeed, the content expertise of social scientists lends itself naturally to 
conversations about the scientific 
implementation of LLMs in research and the societal implications of widespread AI use.

Scholars should keep in mind that LLMs are much more than any one particular use case, and that multiplicity drives both the incredible potential and the fickle and unstable nature of LLM output.
Appropriately incorporating LLMs into the scientific process and building collective knowledge will require both significant humility and creativity, and we believe it will be worth the collective effort.

%
%

\bibliographystyle{unsrtnat}
\bibliography{refs}

\begin{thebibliography}{68}
\providecommand{\natexlab}[1]{#1}
\providecommand{\url}[1]{\texttt{#1}}
\expandafter\ifx\csname urlstyle\endcsname\relax
  \providecommand{\doi}[1]{doi: #1}\else
  \providecommand{\doi}{doi: \begingroup \urlstyle{rm}\Url}\fi

\bibitem[Meincke et~al.(2024)Meincke, Girotra, Nave, Terwiesch, and Ulrich]{meincke2024using}
Lennart Meincke, Karan Girotra, Gideon Nave, Christian Terwiesch, and Karl~T. Ulrich.
\newblock Using large language models for idea generation in innovation.
\newblock \emph{The Wharton School Research Paper Forthcoming}, 9 2024.
\newblock \doi{10.2139/ssrn.4526071}.
\newblock Available at SSRN: https://ssrn.com/abstract=4526071.

\bibitem[Si et~al.(2024)Si, Yang, and Hashimoto]{si2024}
Chenglei Si, Diyi Yang, and Tatsunori Hashimoto.
\newblock Can llms generate novel research ideas? a large-scale human study with 100+ nlp researchers, 2024.
\newblock URL \url{https://doi.org/10.48550/arXiv.2409.04109}.

\bibitem[Schmidgall et~al.(2025)Schmidgall, Su, Wang, Sun, Wu, Yu, Liu, Liu, and Barsoum]{schmidgall2025agentlaboratoryusingllm}
Samuel Schmidgall, Yusheng Su, Ze~Wang, Ximeng Sun, Jialian Wu, Xiaodong Yu, Jiang Liu, Zicheng Liu, and Emad Barsoum.
\newblock Agent laboratory: Using llm agents as research assistants, 2025.
\newblock URL \url{https://arxiv.org/abs/2501.04227}.

\bibitem[Agarwal et~al.(2024)Agarwal, Sahu, Puri, Laradji, Dvijotham, Stanley, Charlin, and Pal]{agarwal2024llms}
Shubham Agarwal, Gaurav Sahu, Abhay Puri, Issam~H Laradji, Krishnamurthy~DJ Dvijotham, Jason Stanley, Laurent Charlin, and Christopher Pal.
\newblock Llms for literature review: Are we there yet?
\newblock \emph{arXiv preprint arXiv:2412.15249}, 2024.

\bibitem[Nejjar et~al.(2025)Nejjar, Zacharias, Stiehle, and Weber]{LLMsForScience}
Mohamed Nejjar, Luca Zacharias, Fabian Stiehle, and Ingo Weber.
\newblock Llms for science: Usage for code generation and data analysis.
\newblock \emph{Journal of Software: Evolution and Process}, 37\penalty0 (1):\penalty0 e2723, 2025.
\newblock \doi{https://doi.org/10.1002/smr.2723}.
\newblock URL \url{https://onlinelibrary.wiley.com/doi/abs/10.1002/smr.2723}.

\bibitem[T{\"o}rnberg(2024)]{tornberg2024large}
Petter T{\"o}rnberg.
\newblock Large language models outperform expert coders and supervised classifiers at annotating political social media messages.
\newblock \emph{Social Science Computer Review}, page 08944393241286471, 2024.

\bibitem[Ornstein et~al.(2023)Ornstein, Blasingame, and Truscott]{ornstein2023train}
Joseph~T Ornstein, Elise~N Blasingame, and Jake~S Truscott.
\newblock How to train your stochastic parrot: Large language models for political texts.
\newblock \emph{Political Science Research and Methods}, pages 1--18, 2023.

\bibitem[Heseltine and Clemm~von Hohenberg(2024)]{heseltine2024large}
Michael Heseltine and Bernhard Clemm~von Hohenberg.
\newblock Large language models as a substitute for human experts in annotating political text.
\newblock \emph{Research \& Politics}, 11\penalty0 (1):\penalty0 20531680241236239, 2024.

\bibitem[Wang(2024)]{wang2024llms}
Yu~Wang.
\newblock Llms in political science: Heralding a new era of visual analysis.
\newblock \emph{arXiv preprint arXiv:2403.00154}, 2024.

\bibitem[Rytting et~al.(2023)Rytting, Sorensen, Argyle, Busby, Gubler, Fulda, and Wingate]{rytting2023}
Christopher Rytting, Taylor Sorensen, Lisa~P. Argyle, Ethan~C. Busby, Joshua~R. Gubler, Nancy Fulda, and David Wingate.
\newblock Towards coding social science datasets with language models.
\newblock \emph{arXiv preprint arXiv:2306.02177}, 2023.
\newblock URL \url{https://arxiv.org/abs/2306.02177}.

\bibitem[Velez and Liu(2023)]{velez2023confronting}
Yamil Velez and Patrick Liu.
\newblock Confronting core issues: A critical test of attitude polarization.
\newblock 2023.

\bibitem[Argyle et~al.(2023)Argyle, Bail, Busby, Gubler, Howe, Rytting, Sorensen, and Wingate]{argyle2023ai}
Lisa~P. Argyle, Christopher~A. Bail, Ethan~C. Busby, Joshua~R. Gubler, Thomas Howe, Christopher Rytting, Taylor Sorensen, and David Wingate.
\newblock Leveraging ai for democratic discourse: Chat interventions can improve online political conversations at scale.
\newblock \emph{Proceedings of the National Academy of the Sciences}, 120\penalty0 (41), 2023.

\bibitem[Tessler et~al.(2024)Tessler, Bakker, Jarrett, Sheahan, Chadwick, Koster, Evans, Campbell-Gillingham, Collins, Parkes, Botvinick, and Summerfield]{tessler2024}
Micahel~Henry Tessler, Michiel~A. Bakker, Daniel Jarrett, Hannah Sheahan, Martin~J. Chadwick, Raphael Koster, Georgina Evans, Lucy Campbell-Gillingham, Tantum Collins, David~C. Parkes, Matthew Botvinick, and Christopher Summerfield.
\newblock Ai can help humans find common ground in democratic deliberation.
\newblock \emph{Science}, 386\penalty0 (6719), 2024.

\bibitem[Hackenburg and Margetts(2024)]{hackenburg_margetts_2023}
Kobi Hackenburg and Helen Margetts.
\newblock Evaluating the persuasive influence of political microtargeting with large language models.
\newblock \emph{Proceedings of the National Academy of Sciences}, 121\penalty0 (24):\penalty0 e2403116121, 2024.

\bibitem[Rozado(2023)]{rozado2023political}
David Rozado.
\newblock The political biases of chatgpt.
\newblock \emph{Social Sciences}, 12\penalty0 (3):\penalty0 148, 2023.

\bibitem[Park et~al.(2023)Park, O'Brien, Cai, Morris, Liang, and Bernstein]{park2023generative}
Joon~Sung Park, Joseph O'Brien, Carrie~Jun Cai, Meredith~Ringel Morris, Percy Liang, and Michael~S Bernstein.
\newblock Generative agents: Interactive simulacra of human behavior.
\newblock In \emph{Proceedings of the 36th annual acm symposium on user interface software and technology}, pages 1--22, 2023.

\bibitem[Palmer and Spirling(2023)]{palmer2023large}
Alexis Palmer and Arthur Spirling.
\newblock Large language models can argue in convincing ways about politics, but humans dislike ai authors: implications for governance.
\newblock \emph{Political science}, 75\penalty0 (3):\penalty0 281--291, 2023.

\bibitem[Argyle and Pope(2022)]{argyle2022does}
Lisa~P Argyle and Jeremy~C Pope.
\newblock Does political participation contribute to polarization in the united states?
\newblock \emph{Public Opinion Quarterly}, 86\penalty0 (3):\penalty0 697--707, 2022.

\bibitem[T{\"o}rnberg et~al.(2023)T{\"o}rnberg, Valeeva, Uitermark, and Bail]{tornberg2023simulating}
Petter T{\"o}rnberg, Diliara Valeeva, Justus Uitermark, and Christopher Bail.
\newblock Simulating social media using large language models to evaluate alternative news feed algorithms.
\newblock \emph{arXiv preprint arXiv:2310.05984}, 2023.

\bibitem[Sreedhar et~al.(2025)Sreedhar, Cai, Ma, Nickerson, and Chilton]{sreedhar2025simulating}
Karthik Sreedhar, Alice Cai, Jenny Ma, Jeffrey~V Nickerson, and Lydia~B Chilton.
\newblock Simulating cooperative prosocial behavior with multi-agent llms: Evidence and mechanisms for ai agents to inform policy decisions.
\newblock In \emph{Proceedings of the 30th International Conference on Intelligent User Interfaces}, pages 1272--1286, 2025.

\bibitem[Bisbee et~al.(2023)Bisbee, Clinton, Dorff, Kenkel, and Larson]{bisbee2023artificially}
James Bisbee, Joshua~D. Clinton, Cassy Dorff, Brenton Kenkel, and Jennifer~M. Larson.
\newblock Synthetic replacements for human survey data? the perils of large language models.
\newblock \emph{Political Analysis}, 32\penalty0 (4):\penalty0 401--416, 2023.

\bibitem[Ashokkumar et~al.(2024)Ashokkumar, Hewitt, Ghezae, and Willer]{ashokkumar2024predicting}
Ashwini Ashokkumar, Luke Hewitt, Isaias Ghezae, and Robb Willer.
\newblock Predicting results of social science experiments using large language models.
\newblock \emph{accessed September}, 19:\penalty0 2024, 2024.

\bibitem[Thorp(2023)]{thorp2023chatgpt}
H~Holden Thorp.
\newblock Chatgpt is fun, but not an author, 2023.

\bibitem[Van~Woudenberg et~al.(2024)Van~Woudenberg, Ranalli, and Bracker]{van2024authorship}
Ren{\'e} Van~Woudenberg, Chris Ranalli, and Daniel Bracker.
\newblock Authorship and chatgpt: A conservative view.
\newblock \emph{Philosophy \& Technology}, 37\penalty0 (1):\penalty0 34, 2024.

\bibitem[Bail(2024)]{bail2024can}
Christopher~A Bail.
\newblock Can generative ai improve social science?
\newblock \emph{Proceedings of the National Academy of Sciences}, 121\penalty0 (21):\penalty0 e2314021121, 2024.

\bibitem[Grossmann et~al.(2023)Grossmann, Feinberg, Parker, Christakis, Tetlock, and Cunningham]{grossmann2023ai}
Igor Grossmann, Matthew Feinberg, Dawn~C Parker, Nicholas~A Christakis, Philip~E Tetlock, and William~A Cunningham.
\newblock Ai and the transformation of social science research.
\newblock \emph{Science}, 380\penalty0 (6650):\penalty0 1108--1109, 2023.

\bibitem[Xu et~al.(2024)Xu, Sun, Ren, Guo, Pan, Lin, Sun, and Han]{xu2024ai}
Ruoxi Xu, Yingfei Sun, Mengjie Ren, Shiguang Guo, Ruotong Pan, Hongyu Lin, Le~Sun, and Xianpei Han.
\newblock Ai for social science and social science of ai: A survey.
\newblock \emph{Information Processing \& Management}, 61\penalty0 (3):\penalty0 103665, 2024.

\bibitem[Argyle et~al.(2025)Argyle, Busby, Gubler, , and Wingate]{argyle2024}
Lisa~P. Argyle, Ethan~C. Busby, Joshua~R. Gubler, , and David Wingate.
\newblock Testing theories of political persuasion using artificial intelligence.
\newblock \emph{Proceedings of the National Academy of the Sciences}, 2025.

\bibitem[Lyman et~al.(2025)Lyman, Hepner, Argyle, Busby, Gubler, and Wingate]{lyman2025smr}
Alex Lyman, Bryce Hepner, Lisa~P. Argyle, Ethan~C. Busby, Joshua~R. Gubler, and David Wingate.
\newblock Balancing large language model alignment and algorithmic fidelity in social science research.
\newblock \emph{Sociological Methods \& Research}, 2025.

\bibitem[McCoy et~al.(2024)McCoy, Yao, Friedman, Hardy, and Griffiths]{mccoy2024embers}
R~Thomas McCoy, Shunyu Yao, Dan Friedman, Mathew~D Hardy, and Thomas~L Griffiths.
\newblock Embers of autoregression show how large language models are shaped by the problem they are trained to solve.
\newblock \emph{Proceedings of the National Academy of Sciences}, 121\penalty0 (41):\penalty0 e2322420121, 2024.

\bibitem[Liu and Shi(2024)]{liu2024poliprompt}
Menglin Liu and Ge~Shi.
\newblock Poliprompt: A high-performance cost-effective llm-based text classification framework for political science.
\newblock \emph{arXiv preprint arXiv:2409.01466}, 2024.

\bibitem[Atreja et~al.(2024)Atreja, Ashkinaze, Li, Mendelsohn, and Hemphill]{atreja2024prompt}
Shubham Atreja, Joshua Ashkinaze, Lingyao Li, Julia Mendelsohn, and Libby Hemphill.
\newblock Prompt design matters for computational social science tasks but in unpredictable ways.
\newblock \emph{arXiv preprint arXiv:2406.11980}, 2024.

\bibitem[Zhuo et~al.(2024)Zhuo, Zhang, Fang, Duan, Lin, and Chen]{zhuo-etal-2024-prosa}
Jingming Zhuo, Songyang Zhang, Xinyu Fang, Haodong Duan, Dahua Lin, and Kai Chen.
\newblock {P}ro{SA}: Assessing and understanding the prompt sensitivity of {LLM}s.
\newblock In \emph{Findings of the Association for Computational Linguistics: EMNLP 2024}. Association for Computational Linguistics, November 2024.
\newblock URL \url{https://aclanthology.org/2024.findings-emnlp.108/}.

\bibitem[Spirling(2023)]{spirling23}
Arthur Spirling.
\newblock Why open-source generative ai models are an ethical way forward for science.
\newblock \emph{Nature}, 616\penalty0 (413), 2023.

\bibitem[Ollion et~al.(2024)Ollion, Shen, Macanovic, and Chatelain]{ollion2024dangers}
{\'E}tienne Ollion, Rubing Shen, Ana Macanovic, and Arnault Chatelain.
\newblock The dangers of using proprietary llms for research.
\newblock \emph{Nature Machine Intelligence}, 6\penalty0 (1):\penalty0 4--5, 2024.

\bibitem[Jaźwińska and Chandrasekar(2025)]{jazwinska2025ai}
Klaudia Jaźwińska and Aisvarya Chandrasekar.
\newblock We compared eight ai search engines. they're all bad at citing news.
\newblock \emph{Columbia Journalism Review}, 2025.

\bibitem[Briggs et~al.(2025)Briggs, Mellon, Arel-Bundock, and Larson]{briggs2025}
Ryan Briggs, Jonathan Mellon, Vincent Arel-Bundock, and Tim Larson.
\newblock We used llms to track methodological and substantive publication patterns in political science and they seem to do a pretty good job, 2025.
\newblock URL \url{https://osf.io/v7fe8}.

\bibitem[Dillion et~al.(2023)Dillion, Tandon, Gu, and Gray]{dillion2023can}
Danica Dillion, Niket Tandon, Yuling Gu, and Kurt Gray.
\newblock Can ai language models replace human participants?
\newblock \emph{Trends in Cognitive Sciences}, 2023.

\bibitem[Velez and Liu(2014)]{velez2024}
Yamil~Ricardo Velez and Patrick Liu.
\newblock Confronting core issues: A critical test of attitude polarization.
\newblock \emph{American Political Science Review}, 2014.

\bibitem[Li et~al.(2023)Li, Hui, Qu, Yang, Li, Li, Wang, Qin, Geng, Huo, Zhou, Chenhao, Li, Chang, Huang, Cheng, and Li]{Li2023}
Jinyang Li, Binyuan Hui, Ge~Qu, Jiaxi Yang, Binhua Li, Bowen Li, Bailin Wang, Bowen Qin, Ruiying Geng, Nan Huo, Xuanhe Zhou, Ma~Chenhao, Guoliang Li, Kevin Chang, Fei Huang, Reynold Cheng, and Yongbin Li.
\newblock How large language models will disrupt data management.
\newblock In \emph{Advances in Neural Information Processing Systems 36 (NeurIPS 2023) Datasets and Benchmarks Track}, 2023.

\bibitem[Fernandez et~al.(2023)Fernandez, Elmore, Franklin, Krishnan, and Tan]{Fernandez2023}
Raul~Castro Fernandez, Aaron~J. Elmore, Michael~J. Franklin, Sanjay Krishnan, and Chenhao Tan.
\newblock How large language models will disrupt data management.
\newblock In \emph{Proceedings of the VLDB Endowment}, volume~16, pages 3302--3309, July 2023.
\newblock \doi{https://doi.org/10.14778/3611479.3611527}.

\bibitem[Xiao et~al.(2023)Xiao, Xu, Zhang, Wang, and Xia]{Xiao2023}
Changrong Xiao, Sean~Xin Xu, Kunpeng Zhang, Yufang Wang, and Lei Xia.
\newblock Evaluating reading comprehension exercises generated by llms: A showcase of chatgpt in education applications.
\newblock In \emph{Proceedings of the 18th Workshop on Innovative Use of NLP for Building Educational Applications (BEA 2023)}, pages 610--625, July 2023.
\newblock \doi{https://aclanthology.org/2023.bea-1.52/}.

\bibitem[Lyu et~al.(2024)Lyu, Wang, Chung, Sun, and Zhang]{lyu2024}
Wenhan Lyu, Yimeng Wang, Tingting~(Rachel) Chung, Yifan Sun, and Yixuan Zhang.
\newblock Evaluating the effectiveness of llms in introductory computer science education: A semester-long field study.
\newblock In \emph{L@S '24: Proceedings of the Eleventh ACM Conference on Learning @ Scale}, pages 63--74, July 2024.
\newblock \doi{https://doi.org/10.1145/3657604.3662036}.

\bibitem[Milano et~al.(2023)Milano, McGrane, and Leonelli]{milano2023}
Silvia Milano, Joshua~A. McGrane, and Sabina Leonelli.
\newblock Large language models challenge the future of higher education.
\newblock \emph{Nature Machine Intelligence}, 5:\penalty0 333--334, 2023.

\bibitem[Yakura et~al.(2024)Yakura, Lopez-Lopez, Brinkmann, Serna, Gupta, and Rahwan]{yakura2024empiricalevidencelargelanguage}
Hiromu Yakura, Ezequiel Lopez-Lopez, Levin Brinkmann, Ignacio Serna, Prateek Gupta, and Iyad Rahwan.
\newblock Empirical evidence of large language model's influence on human spoken communication, 2024.
\newblock URL \url{https://arxiv.org/abs/2409.01754}.

\bibitem[Hohenstein et~al.(2023)Hohenstein, Kizilcec, DiFranzo, Aghajari, Mieczkowski, Levy, Naaman, Hancock, and Jung]{Hohenstein2023}
Jess Hohenstein, René~F. Kizilcec, Dominic DiFranzo, Zhila Aghajari, Hannah Mieczkowski, Karen Levy, Mor Naaman, Jeffrey Hancock, and Malte~F. Jung.
\newblock Artificial intelligence in communication impacts language and social relationships.
\newblock \emph{Scientific Reports}, 13\penalty0 (1):\penalty0 5487, 2023.
\newblock ISSN 2045-2322.
\newblock \doi{10.1038/s41598-023-30938-9}.
\newblock URL \url{https://doi.org/10.1038/s41598-023-30938-9}.

\bibitem[King et~al.(2021)King, Keohane, and Verba]{kkv2021}
Gary King, Robert~O. Keohane, and Sidney Verba.
\newblock \emph{Designing Social Inquiry: Scientific Inference in Qualitative Research, new edition}.
\newblock Princeton University Press, Chicago, 2021.

\bibitem[Lakatos(1970)]{lakatos1970_2}
Imre Lakatos.
\newblock Falsification and the methodology of scientific research programmes.
\newblock In Imre Lakatos and Alan Musgrave, editors, \emph{Criticism and the Growth of Knowledge: Proceedings of the International Colloquium in the Philosophy of Science, London, 1965}, pages 91--196, New York, 1970. Cambridge University Press.

\bibitem[Cui et~al.(2024{\natexlab{a}})Cui, Li, and Zhou]{Cui2024}
Ziyan Cui, Ning Li, and Huaikang Zhou.
\newblock Can ai replace human subjects? a large-scale replication of psychological experiments with llms, 2024{\natexlab{a}}.
\newblock URL \url{https://doi.org/10.48550/arXiv.2409.00128}.

\bibitem[Horton(2023)]{horton2023large}
John~J Horton.
\newblock Large language models as simulated economic agents: What can we learn from homo silicus?
\newblock Technical report, National Bureau of Economic Research, 2023.

\bibitem[Lippert et~al.(2024)Lippert, Dreber, Johannesson, Tierney, Cyrus-Lai, Uhlmann, Collaboration, and Pfeiffer]{Lippert2024}
Steffen Lippert, Anna Dreber, Magnus Johannesson, Warren Tierney, Wilson Cyrus-Lai, Eric~Luis Uhlmann, Emotion~Expression Collaboration, and Thomas Pfeiffer.
\newblock Can large language models help predict results from a complex behavioural science study?
\newblock \emph{Royal Society Open Science}, 11:\penalty0 240682, 2024.

\bibitem[Hackenburg et~al.(2025)Hackenburg, Tappin, R{\"o}ttger, Hale, Bright, and Margetts]{hackenburg2025scaling}
Kobi Hackenburg, Ben~M Tappin, Paul R{\"o}ttger, Scott~A Hale, Jonathan Bright, and Helen Margetts.
\newblock Scaling language model size yields diminishing returns for single-message political persuasion.
\newblock \emph{Proceedings of the National Academy of Sciences}, 122\penalty0 (10):\penalty0 e2413443122, 2025.

\bibitem[Szymanski et~al.(2025)Szymanski, Ziems, Eicher-Miller, Li, Jiang, and Metoyer]{szymanski2025limitations}
Annalisa Szymanski, Noah Ziems, Heather~A Eicher-Miller, Toby Jia-Jun Li, Meng Jiang, and Ronald~A Metoyer.
\newblock Limitations of the llm-as-a-judge approach for evaluating llm outputs in expert knowledge tasks.
\newblock In \emph{Proceedings of the 30th International Conference on Intelligent User Interfaces}, pages 952--966, 2025.

\bibitem[Huang et~al.(2024)Huang, Lam, Li, Ren, Wang, Jiao, Tu, and Lyu]{huang2024apathetic}
Jen-tse Huang, Man~Ho Lam, Eric~John Li, Shujie Ren, Wenxuan Wang, Wenxiang Jiao, Zhaopeng Tu, and Michael~R Lyu.
\newblock Apathetic or empathetic? evaluating llms' emotional alignments with humans.
\newblock \emph{Advances in Neural Information Processing Systems}, 37:\penalty0 97053--97087, 2024.

\bibitem[Amirizaniani et~al.(2024)Amirizaniani, Martin, Sivachenko, Mashhadi, and Shah]{amirizaniani2024can}
Maryam Amirizaniani, Elias Martin, Maryna Sivachenko, Afra Mashhadi, and Chirag Shah.
\newblock Can llms reason like humans? assessing theory of mind reasoning in llms for open-ended questions.
\newblock In \emph{Proceedings of the 33rd ACM International Conference on Information and Knowledge Management}, pages 34--44, 2024.

\bibitem[Valmeekam et~al.(2022)Valmeekam, Olmo, Sreedharan, and Kambhampati]{valmeekam2022large}
Karthik Valmeekam, Alberto Olmo, Sarath Sreedharan, and Subbarao Kambhampati.
\newblock Large language models still can't plan (a benchmark for llms on planning and reasoning about change).
\newblock In \emph{NeurIPS 2022 Foundation Models for Decision Making Workshop}, 2022.

\bibitem[Eaton(2024)]{eaton2024}
Kim Eaton.
\newblock How many r’s in ‘strawberry’? this ai doesn’t know, 2024.
\newblock URL \url{https://www.inc.com/kit-eaton/how-many-rs-in-strawberry-this-ai-cant-tell-you.html}.

\bibitem[Lu et~al.(2024)Lu, Zhu, Li, Qiao, and Yuan]{lu2024llamax}
Yinquan Lu, Wenhao Zhu, Lei Li, Yu~Qiao, and Fei Yuan.
\newblock Llamax: Scaling linguistic horizons of llm by enhancing translation capabilities beyond 100 languages.
\newblock \emph{arXiv preprint arXiv:2407.05975}, 2024.

\bibitem[Kaplan et~al.(2020)Kaplan, McCandlish, Henighan, Brown, Chess, Child, Gray, Radford, Wu, and Amodei]{kaplan2020scalinglawsneurallanguage}
Jared Kaplan, Sam McCandlish, Tom Henighan, Tom~B. Brown, Benjamin Chess, Rewon Child, Scott Gray, Alec Radford, Jeffrey Wu, and Dario Amodei.
\newblock Scaling laws for neural language models, 2020.
\newblock URL \url{https://arxiv.org/abs/2001.08361}.

\bibitem[Cui et~al.(2024{\natexlab{b}})Cui, Chiang, Stoica, and Hsieh]{cui2024orbenchoverrefusalbenchmarklarge}
Justin Cui, Wei-Lin Chiang, Ion Stoica, and Cho-Jui Hsieh.
\newblock Or-bench: An over-refusal benchmark for large language models, 2024{\natexlab{b}}.
\newblock URL \url{https://arxiv.org/abs/2405.20947}.

\bibitem[Tekgurler(2025)]{tekgurler2025llms}
Merve Tekgurler.
\newblock Llms for translation: Historical, low-resourced languages and contemporary ai models.
\newblock \emph{arXiv preprint arXiv:2503.11898}, 2025.

\bibitem[Kirk et~al.(2024)Kirk, Mediratta, Nalmpantis, Luketina, Hambro, Grefenstette, and Raileanu]{kirk2024understanding}
Robert Kirk, Ishita Mediratta, Christoforos Nalmpantis, Jelena Luketina, Eric Hambro, Edward Grefenstette, and Roberta Raileanu.
\newblock Understanding the effects of {RLHF} on {LLM} generalisation and diversity.
\newblock In \emph{The Twelfth International Conference on Learning Representations}, 2024.
\newblock URL \url{https://openreview.net/forum?id=PXD3FAVHJT}.

\bibitem[Li et~al.(2024)Li, Ding, Fang, and Tao]{li-etal-2024-revisiting}
Hongyu Li, Liang Ding, Meng Fang, and Dacheng Tao.
\newblock Revisiting catastrophic forgetting in large language model tuning.
\newblock In Yaser Al-Onaizan, Mohit Bansal, and Yun-Nung Chen, editors, \emph{Findings of the Association for Computational Linguistics: EMNLP 2024}, pages 4297--4308, Miami, Florida, USA, November 2024. Association for Computational Linguistics.
\newblock \doi{10.18653/v1/2024.findings-emnlp.249}.

\bibitem[Pezeshkpour and Hruschka(2024)]{pezeshkpour-hruschka-2024-large}
Pouya Pezeshkpour and Estevam Hruschka.
\newblock Large language models sensitivity to the order of options in multiple-choice questions.
\newblock In Kevin Duh, Helena Gomez, and Steven Bethard, editors, \emph{Findings of the Association for Computational Linguistics: NAACL 2024}, pages 2006--2017, Mexico City, Mexico, June 2024. Association for Computational Linguistics.
\newblock \doi{10.18653/v1/2024.findings-naacl.130}.

\bibitem[Wang et~al.(2025)Wang, Wu, Tang, Luo, Chen, Chen, and He]{wang2025limits}
Qian Wang, Jiaying Wu, Zhenheng Tang, Bingqiao Luo, Nuo Chen, Wei Chen, and Bingsheng He.
\newblock What limits llm-based human simulation: Llms or our design?
\newblock \emph{arXiv preprint arXiv:2501.08579}, 2025.

\bibitem[Santurkar et~al.(2023)Santurkar, Durmus, Ladhak, Lee, Liang, and Hashimoto]{santurkar2023whose}
Shibani Santurkar, Esin Durmus, Faisal Ladhak, Cinoo Lee, Percy Liang, and Tatsunori Hashimoto.
\newblock Whose opinions do language models reflect?
\newblock In \emph{Proceedings of the 40th International Conference on Machine Learning}, pages 1--34, 2023.

\bibitem[Boelaert et~al.(2024)Boelaert, Coavoux, Ollion, Petev, and Pr{\"a}g]{boelaert2024machine}
Julien Boelaert, Samuel Coavoux, {\'E}tienne Ollion, Ivaylo Petev, and Patrick Pr{\"a}g.
\newblock Machine bias generative large language models have a worldview of their own.
\newblock 2024.

\bibitem[Kim and Lee(2023)]{kim2023ai}
Junsol Kim and Byungkyu Lee.
\newblock Ai-augmented surveys: Leveraging large language models and surveys for opinion prediction.
\newblock \emph{arXiv preprint arXiv:2305.09620}, 2023.

\end{thebibliography}

\end{document}